# A reentrant phenomenon in magnetic and dielectric properties of $Dy_2BaNiO_5$ and an intriguing influence of external magnetic field

Tathamay Basu,[1] P.L. Paulose,[1] Kartik K Iyer,[1] Kiran Singh,[1,#] N. Mohapatra,[2] S. Chowki,[2] Babu Gonde[1] and E.V. Sampathkumaran[1,*]

[1]*Tata Institute of Fundamental Research, Homi Bhabha Road, Colaba, Mumbai 400005, India*
[2]*School of Basic Sciences, Indian Institute of Technology - Bhubaneswar, Bhubaneswar 751013, India*

We report that the spin-chain compound $Dy_2BaNiO_5$, recently proven to exhibit magnetoelectric coupling below its Néel temperature ($T_N$) of 58 K, exhibits strong frequency-dependent behavior in ac magnetic susceptibility and complex dielectric properties at low temperatures (<10 K), mimicking 'reentrant' multiglass phenomenon. Such a behavior is not known among undoped compounds. A new finding in the field of multiferroics is that the characteristic magnetic feature at such low temperatures moves towards higher temperatures in the presence of a magnetic-field ($H$), whereas the corresponding dielectric feature shifts towards lower temperatures with $H$, unlike the situation near $T_N$. This observation indicates that the alignment of spins by external magnetic fields tends to inhibit glassy-like slow electric-dipole dynamics, at least in this system, possibly arising from peculiarities in the magnetic structure.

**J. Phys.: Condensed Matter (Fast Track Communication), in press**

In the field of 'multiferroics' dealing with studies of coupled ferroelectric and magnetic ordering [see, for instance, Ref. 1-4], very little knowledge exists on how such magnetoelectrically (ME) coupled transitions respond to external parameters like magnetic-field ($H$) in the event that these ME phenomena are glassy-like in their behavior induced by disorder or geometric factors, which cannot be ruled out in nanometric length scales even in bulk form of materials. It is of academic interest to understand various factors contributing to such responses apart from its relevance for any practical application. The phenomenon of spin-glass, originally observed in dilute magnetic systems, is abundant in concentrated magnetic systems [5] and the concept of cooperative 'dipolar-glass' ordering has been known in systems with dipole impurities [6]. However, the concept of 'multiglass', that is, the coexistence of spin-glass and dipolar glass phenomena in the field of multiferroics was not realized till Kleemann et al [7,8] proposed it for a few chemically-doped disordered systems, following which such a phenomenon has been reported in a few other systems [9-13]. Such reports, particularly on undoped materials, are still warranted for the advancement of knowledge in this field. In this Fast Track Communication, we report a 'reentrant glassy-like behavior' in *ac* magnetic susceptibility ($\chi$) and complex dielectric properties at temperatures far below the Néel temperature ($T_N$= 58 K), as inferred from the strong frequency ($\nu$) dependence of these properties, in a Haldane spin-chain compound, $Dy_2BaNiO_5$, crystallizing in an orthorhombic structure (space group: *Immm*) [14]. It is intriguing to note that an application of external magnetic field influences these glassy-like magnetic and electric dipole anomalies differently.

This rare-earth family attracted considerable attention from various angles [see, for instance, Refs. 15-16] as briefed in Ref. 17. Recently, for $Dy_2BaNiO_5$, we reported [17] subtle and broad magnetic anomalies around 10 K and 30-50 K, apart from the one at $T_N$, in heat-capacity ($C$), *ac* and *dc* $\chi$ and we attributed these to gradual changes in the relative orientations of the magnetic moments of Dy and Ni. Interestingly, we observed magnetoelectric coupling in these temperature ranges, with a possible evidence for multiferroic behavior [17]. An external magnetic field should significantly influence the relative orientation of the Dy and Ni moments; as a consequence, the magnetic and dielectric features as a function of temperature can be modified. This prompted us to carry out detailed magnetic and dielectric studies on this material.

The polycrystalline samples employed in the present investigation and the equipments used to measure $T$-dependence of *ac* $\chi$ and magnetization, complex dielectric behavior and heat-capacity are the same as those in previous studies [17]. *Ac* $\chi$ measurements were performed with the help of a commercial Physical Properties Measurement System (Quantum Design (QD), USA) with an ac field of 1 Oe in the frequency range 13 Hz to 10 kHz in fixed *dc* external magnetic fields (0, 50 and 90 kOe) and the complex dielectric permittivity was measured with 1 V ac bias during warming (with a rate of 1K/min) in the frequency range 1 - 30 kHz in various external fields up to 140 kOe. Additional *ac* $\chi$ measurements below 1.3 kHz were carried out by a Superconducting Quantum Interference Device (SQUID, QD) magnetometer. $C(T)$ behavior was also tracked in the presence several magnetic fields up to 140 kOe. All the measurements were performed for zero-field-cooled condition (from 150 K) of the specimen and the results as obtained on different specimens are found to be robust.

The results of *ac* $\chi$ measurements are shown in figure 1. In this paragraph, we discuss the data obtained in the absence of an external magnetic-field. The shape of the curve obtained for the real part ($\chi'$) (see the inset of figure 1) in the absence of magnetic-field is similar to that reported in *dc* $\chi$ [17]. The feature at $T_N$ is not well-resolved; there is a broad peak around 40 K followed by a fall with lowering temperature attributable to the opening of the well-known spin-gap; presumably a gradual change in the ordered Dy moment and relative orientations of magnetic moments of Dy and Ni may also contribute to this [16]. These features well above 10 K are $\nu$-independent and imaginary part ($\chi''$) is also featureless, thereby establishing that the magnetism is not of a glassy-type in this temperature range. We do not elaborate this feature any further. We show the data below 40 K in the mainframe of figure 1 to highlight low temperature anomalies. There is an upturn in $\chi'$ below about 10 K as though there is a peak at temperatures below 2 K. We have actually measured *ac* $\chi$ even with a frequency of 1.3 Hz with SQUID (but not shown for the sake of clarity of the figure) and it was noted that $\chi''$ varies negligibly down to 1.8K, but an increase in $\nu$ to 13 Hz reveals a low-temperature upturn in both the components, eventually resulting in clear peaks in $\chi''$ for higher values of $\nu$ (see the curve for 10 kHz). The $\chi'$ curves obtained at various frequencies are well-separated below 12 K, which can be correlated to the onset of a magnetic anomaly below this temperature as discussed in Ref. 17. Judged by the behavior of the peak in $\chi''(T)$, one can safely state that this magnetic feature is strongly frequency-dependent (about 6 K for a variation of $\nu$ from 13 Hz to 10 kHz). Though this change in peak-temperature for a variation of frequency is quite large compared to that seen in canonical spin-glasses [5], such a behavior has been reported to arise from complex spin dynamics in geometrically frustrated magnetic systems, e.g., $Ca_3Co_2O_6$ [18-21]. We believe that that the low temperature feature (below 10 K) is intrinsic bulk effect and not due to grain boundary effects, considering that the features are not weak, particularly in heat-capacity data (see below).

We have performed isothermal remnant magnetization ($M_{IRM}$) measurements at various temperatures. For this purpose, the specimen was cooled in zero field from the paramagnetic state to desired temperatures. Then a magnetic-field of 5 kOe was turned on for 5 mins and $M_{IRM}$ was measured after switching off the field as a function of time (*t*) for more than an hour. We have observed a qualitative change in the behavior of $M_{IRM}$ with temperature in the sense that there is a very slow decay of $M_{IRM}$ at very low temperatures (see, for instance, the curves in the 2 - 5 K range in figure 2, top panel) and the time over which the decay persists decreases with increasing temperature. It is important to note that, at 12 K, the decay gets sharper with a qualitative change in the shape of the plot (and at 15 K the decay occurred within few mins). This observation, apart from revealing that there is definitely a change in the magnetic character of the material around 10 - 12 K, provides compelling evidence for the glassy nature of the magnetism at such low temperatures. We have also attempted to fit the $M_{IRM}$ curves to a stretched exponential form [22] and we are able to fit well for the curves below 5 K, but not for higher temperatures as seen in the bottom panel of figure 2. The relaxation time ($\tau$) for the magnetic process obtained appears to fall in a narrow range 200 - 700 s with a scatter in the values and therefore it is rather difficult to get any meaningful information about the validity of Arrhenius behavior as inferred from the plot of $\tau$ versus $1/T$ (see inset of figure 2).

We would like to mention that we searched for an additional signature for spin-glasses, namely 'memory effects' [8]. For this purpose, we obtained *dc* $\chi$ data down to 2 K after waiting for several hours (3 h, 6 h) at different temperatures below 10 K and without waiting. We could not resolve any 'local dip' at the waiting temperatures in the difference curves. It is therefore difficult to claim that the low-temperature (<10K) magnetic state can be described as a conventional spin-glass.

We now turn to the complex dielectric behavior. Figure 3a shows *T*-dependence of dielectric constant ($\varepsilon'$) as well as loss factor (*tanδ*) obtained with various frequencies (in the absence of an external magnetic-field). The derivative curves revealed evidence for magnetoelectric coupling at $T_N$ and below $T_N$ in our earlier publication and therefore we will not elaborate this aspect here, except to recall that this material is highly insulating and that the observed magnetoelectric coupling is intrinsic [17]. A new observation we are bringing out here is that, well below $T_N$, an increase in frequency from 1 kHz to 30 kHz causes a significant shift of the curves to higher temperature ranges. We focus on two temperature ranges: i) In the range 20 - 40 K, the curve shifts as much as about 10 K, which is more transparent from the movement of the 20-30 K peak in *tanδ* (see inset of figure 3 for zero field behavior). This trend qualitatively persists even for high fields (not shown).

Though one can not pinpoint a particular temperature for a well-defined magnetic transition, there is a gradual change of the magnetic moment of Dy as well as in its angle with Ni magnetic moment [16]; it is not clear whether this magnetic process results in such a frequency dependence of complex dielectric permittivity, despite the absence of glassy magnetic behavior in this *T*-range. ii) There is another well-defined feature (a change of slope) around 10 K. A point of emphasis is that this 10K-feature, which is more well-defined compared to the one at the higher temperature range, also exhibits finite frequency dependence. Thus, for instance, the temperature at which ε' falls with lowering temperature shifts from 10 K to 14-16 K when ν is increased from 1 kHz to 30 kHz. There are corresponding changes in *tanδ* as well. Thus, these dielectric data reveal an important characteristic feature of dielectric glasses [23]. However, as in the case of dc χ, we could not resolve the corresponding feature in 'memory' experiments. However, considering that this is observed in the same temperature range in which such glassy features are seen in ac χ, we conclude that this glassy-like electric dipolar state is also due to ME coupling. Incidentally, we note that the temperature at which *tanδ(T)* peaks for a given frequency is found to follow Arrhenius behavior (as shown in the inset of figure 3b) and activation energy for this process turns out to be about 85 K with apparently negligible change for an application of 50 kOe. This observation is different from that observed for the magnetic process, thereby revealing complexities involved in these coupled properties. We would also like to add that we have obtained Cole-Cole plot (that is, the plot of real part of impedance versus its imaginary part) at different bias voltages (0.5 - 2V) at 3 K and we found that these plots are not semicircles and also do not exhibit any bias-voltage dependence, as observed at 50 K remarked earlier [17]. Therefore, we conclude that the observed behaviour is due to grain rather than grain boundary and hence the grain boundary (and other extrinsic) effects can be ignored.

Clearly, the scenario at low temperatures mimics 'reentrant phenomenon' - unusual among undoped stoichiometric compounds in the field of multiferroics - considering that multiferroic behavior due to antiferromagnetic ordering is observed at a higher *T*-range. However, we are hesitant to classify the low-temperature state into 'multiglass' of the type discussed in Refs. 7-12, primarily because we are not able to resolve a characteristic 'local dip' in 'memory' experiments. We believe that slow and complex dynamics, similar to the case of $Ca_3Co_2O_6$ (see, for instance, Ref. 24 and article cited therein), arising from changes in the relative orientations of antiferromagnetically coupled Dy and Ni moments, result in glass-like anomalies depending on the time-scale of measurement. In fact, even for the case of $Ca_3Co_2O_6$, we have not observed the 'local dip' at the waiting temperature in 'memory' experiments (unpublished), supporting our belief that the apparent 'multi-glassiness' observed in some measurements in such materials may have a different origin with respect to that in other systems [7-12]. It is also possible that structural disorder plays a role. In the material under study, though the magnetic structure at 1.5 K has been classified to be of an antiferromagnetic type with Dy and Ni moments lying essentially along *c*-direction [16], it is possible that the magnetic chains are also broken due to disorder and the glassy anomalies can also arise from selected clusters with a possible spread in characteristic glassy temperatures and spin-dynamics depending on the size of the clusters.

Now, we present the results of our investigation how an application of external magnetic-field influences the features. In order to address how the magnetism is influenced, we first focus on *C(T)/T* curves in the presence of external fields. We have noted (see figure 4a) that there is a gradual reduction in the intensity and smearing of the peak at $T_N$, but it appears that there is negligible suppression of the value of $T_N$ with increasing *H*. At intermediate temperatures (~ 15 - 40 K), in the zero-field data, there is a convex-shaped curvature which is suppressed gradually by the external field (see the inset of figure 4a, shown for some magnetic fields) and it is not straightforward to draw any conclusion based on this observation. Focussing at a further low temperature range (< 15 K) (see figure 4b), in zero field, there is a change of slope around 10 K, followed by an upturn below 7 K (λ-anomaly); this appears to be the result of a magnetic feature discussed in Ref. 17. This upturn is gradually suppressed by the magnetic-field and a clear peak is observed at 8 K with the minimum around 10 K for 50 kOe. For 70 kOe (after the metamagnetic transition at about 60 kOe [17]), these features move to higher temperatures of 12 and 14 K respectively. Thereafter these temperatures are essentially unaffected for higher values of *H*. These findings suggest that the external fields tend to shift these magnetic features to higher temperature range.

We have also performed *ac* $\chi$ measurements in the presence of 50 and 90 kOe (see figure 1). Both $\chi'$ and $\chi''$ curves show clear peaks in the same *T*-range in which *C(T)* exhibits a $\lambda$-anomaly in the vicinity of 10 K, even at low frequencies unlike the zero-field data. In particular, the cusp which was not seen above 2 K in $\chi'$ is shifted to higher temperatures at high magnetic fields, similar to the behavior observed in *C(T)*. There is still a finite frequency-dependence of the peak (see $\chi'$ plots) for *H*= 50 kOe (about 4 K as $\nu$ is varied from 13 Hz to 10 kHz) at the verge of metamagnetic transition. As seen in $\chi'$ plots, the $\nu$-dependence is relatively suppressed for *H*= 90 kOe which is a measure of reduced glassiness. It may be stated that the curves of $\chi$(T) (particularly in the vicinity of the peak) at high fields attain dramatically higher values with respect to those of zero-field and there is a non-monotonic variation of the intensities of the peaks when *H* is increased. It is not clear whether complex nature of the magnetism and possible *H*-dependence of spin-gap are responsible for this. The capabilities of our instrument do not permit us to see how these features get modified at much higher fields.

In figure 5, we show the complex dielectric permittivity data obtained in the presence of external magnetic-fields with a frequency of 30 kHz. There is no appreciable shift in the peak temperature of the derivative of *tan$\delta$* near $T_N$ with increasing magnetic field (see figure 5a), tracking the inference from *C(T)* data described above (Fig. 4a). In contrast to this, surprisingly, the dielectric features, $\varepsilon'$ and *tan $\delta$*, associated with the 10K-feature shift towards low-*T* region with increasing *H (*see figure 5b). Thus, the change of slope visible in zero field in the plot of $\varepsilon'$(*T*) for this frequency near 15 K gets more prominent in the form of a peak for higher values of *H* with this characteristic temperature getting lowered. For instance, the peak in $\varepsilon'$(*T*) appears at 12 and 10 K for *H*= 50 and 140 kOe respectively. This behavior is in sharp contrast to those of ac $\chi$ and heat capacity features. Similar trends are observed in the plots for all other frequencies employed, though the corresponding peak temperatures are different (see Fig. 3) for different frequencies. It is worth noting that, as in the case of *ac $\chi$* (see figure 1), the peak intensities in $\varepsilon'$(*T*) and *tan$\delta$(T)* increase with *H* initially (up to 50 kOe in figure 5), but decreases for higher values of *H*, thereby supporting that there is a coupling between magnetic component and electric dipoles.

In short, the temperatures characterizing these two coupled phenomena near 10 K respond differently to applications of magnetic-fields, strikingly without any such anomaly at *$T_N$*, as elaborated above. This finding is quite intriguing, and we speculate that this apparent 'disassociation' below 10 K results from the spin alignment by an external magnetic field, which in turn seems to disfavour 'glassy' electric dipole ordering. We think that this 'dissociation' can also be considered as a manifestation of ME coupling, as otherwise one would naively expect that the characteristic temperature (near 10 K) in complex dielectric permittivity curves does not move at all with the application of magnetic-fields. A qualitative comparison of the trend in the peak intensities in figure 1 and figure 5 (as mentioned earlier) also supports the persistence of some kind of coupling even with the application of magnetic fields. As a possible manifestation of this dissociation, at high fields, the $\nu$-dependence of dielectric constant is also reduced as evident from figures 3b and 3c. Careful neutron diffraction measurements as a function of magnetic-field are warranted to understand the degree of randomness of magnetic moments in zero field and its suppression by external field to test our hypothesis. At this juncture, we would like to stress that we carried out similar investigations on the Er member (for which $T_N$ is ~32 K) and we found a 'reentrant behavior' below about 6 K even in this case. However, we noted that the response of the features due to magnetism and electric dipoles to the application of the magnetic-fields are the same in the Er case (unpublished). It is known [16] that orbital anisotropy plays a major role in deciding the magnetic behavior, resulting in a difference in the alignments of magnetic moments of Dy and Er members. We therefore wonder whether this anisotropy also plays a role on the observed $H$-dependent anomaly seen for the Dy sample. Incidentally, in support of this proposal, we did not see these glassy features at all in the analogous Gd (S-state ion) member ($T_N$= 58 K) down to 2 K (unpublished). Clearly, the dichotomy of magnetic and dielectric behavior of the present compound raises interesting questions.

To conclude, the compound $Dy_2BaNiO_5$ is found to exhibit a strong frequency dependence of ac magnetic susceptibility and dielectric constant in a temperature range far below $T_N$, typical of that expected for a 'reentrant multiglass', thereby demonstrating interesting slow-dynamics of spin and electric dipoles in this material at such low temperatures. Interestingly, an application of magnetic field has an opposite effect on magnetic and electric dipole ordering at low temperatures well below $T_N$, and it appears that the peculiarities associated with anisotropic magnetic structure in this family crucially play a role on this opposite trend.

**References**

#Current address: UGC-DAE Consortium for Scientific Research, University Campus, Khandwa Road, Indore - 452001, India

[1] Kimura T, Goto T, Shintani H, Ishizaka K, Arima T, and Tokura Y 2003 Nature (London) **426** 55; Kimura T, Seiko Y, Nakamura H, Seigrist T and Ramirez A P 2008 Nature Mater. **7** 291
[2] Subramanian M A, He T, Chain J, Rogado N S, Calvarese T G and Sleight A W 2006 Adv. Mater. **18** 1737
[3] Ko K T, Jung M H, He Q, Lee J H and Woo C S 2001 Nat. Commun. **2** 567
[4] Rao C N R, Sundaresan A, and Saha R 2012 J. Phys. Chem. Lett. **3** 2237
[5] Binder K and Young A P 1986 Rev. Mod. Phys. **58,** 801
[6] Vugmeister B E and Glinchuk M D 1990 Rev. Mod. Phys. **62,** 993
[7] Kleemann W and Shvartsman V V, Borisov P, and Kania A 2010 Phys. Rev. Lett. **105,** 257202
[8] Kleemann W, 2012 Sol. State Phen. **189,** 41
[9] Shvartsmnn, V V, Bedanta S, Borisov P, Kleeman W, Tkach A and Vilarinho P M 2008 *Phys. Rev. Lett*. **101** 165704
[10] Choudhury D D, Mandal P, Mathieu R, Hazarika A, Rajan, S, Sundaresan A, Waghmare U V, Knut R, Karis O, Nordblad P and Sarma D D 2012 Phys. Rev. Lett. **108**, 127201
[11] Singh K, Maignan A, Simon C, Kumar S, Martin C, Lebedev O, Turner S and Tendeloo G V J. Phys.: Condens. Matter. 2012 **24** 226002
[12] Yamaguchi Y, Nakano T, Nozue Y and Kimura T 2012 Phys. Rev. Lett. **108** 057203
[13] Basu T, Iyer K K, Singh K, and Sampathkumaran E V 2013 Sci. Rep. **3,** 3104
[14] Darriet J and Regnault L P 1993 Solid State Commun. **86,** 409.
[15] Yokoo T, Zheludev A, Nakamura M and Akimitsu J 1997 Phys. Rev. B **55** 11516
[16] Garcia-Matres E, Martinez J L and Rodriguez-Carvajal J 2001 Eur. Phys. J. B. **24** 59
[17] Singh K, Basu T, Chowki S, Mahapotra N, Iyer K, Paulose P L and Sampathkumaran E V 2013 Phys. Rev. B **88** 094438
[18] Rayaprol S, Sengupta K and Sampathkumaran E V 2003 Solid State Commun. **128** 79
[19] Maignan A, Michel C, Masset AC, Martin C and Raveau B 2000 Eur. Phys. J. B **15** 657
[20] Sampathkumaran E V and Niazi A 2002 Phys. Rev. B **65** 180401(R).
[21] Rayaprol S, Sengupta K and E.V. Sampathkumaran 2003 Phys. Rev. B **67** 180404(R)
[22] Suzuki I S and Suzuki M 2008 Phys. Rev. B **78** 214404.

[23] Bhattacharya S, Nagel S R, Fleishman L and Susman S 1982 *Phys. Rev. Lett.* **48** 1267

[24] Paddison J A M, arXiv:cond-mat/1312.5243 (2013).

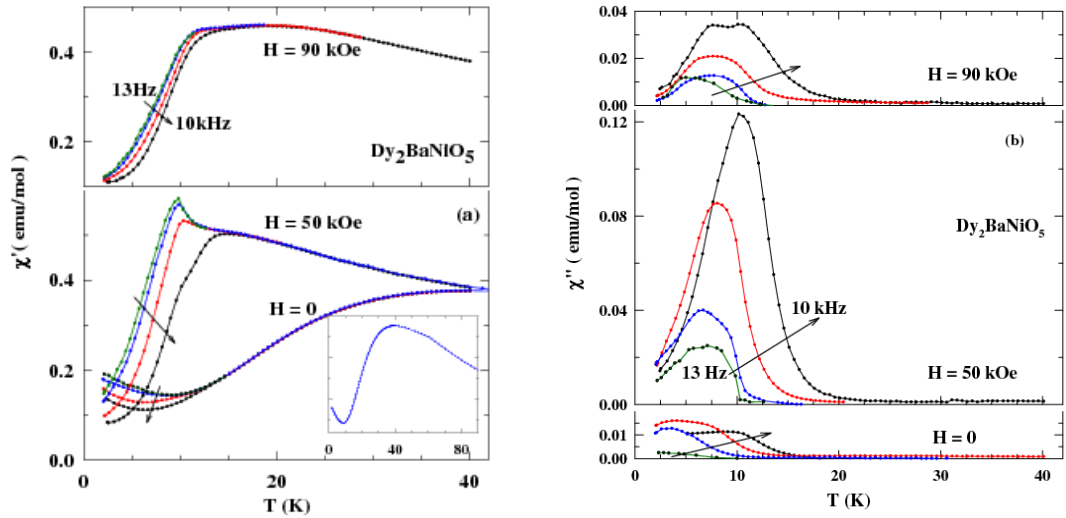

Figure 1:
(a) Real ($\chi'$) and (b) imaginary ($\chi''$) parts of ac magnetic susceptibility measured with various frequencies (13 Hz, 130 Hz, 1.3 kHz and 10 kHz) in the absence and in the presence of magnetic fields, as a function of temperature below 40 K. In the inset, the data (measured in the zero field) over a wider temperature range (below 80 K) for 130 Hz is plotted. The lines through the data points serve as guides to eyes.

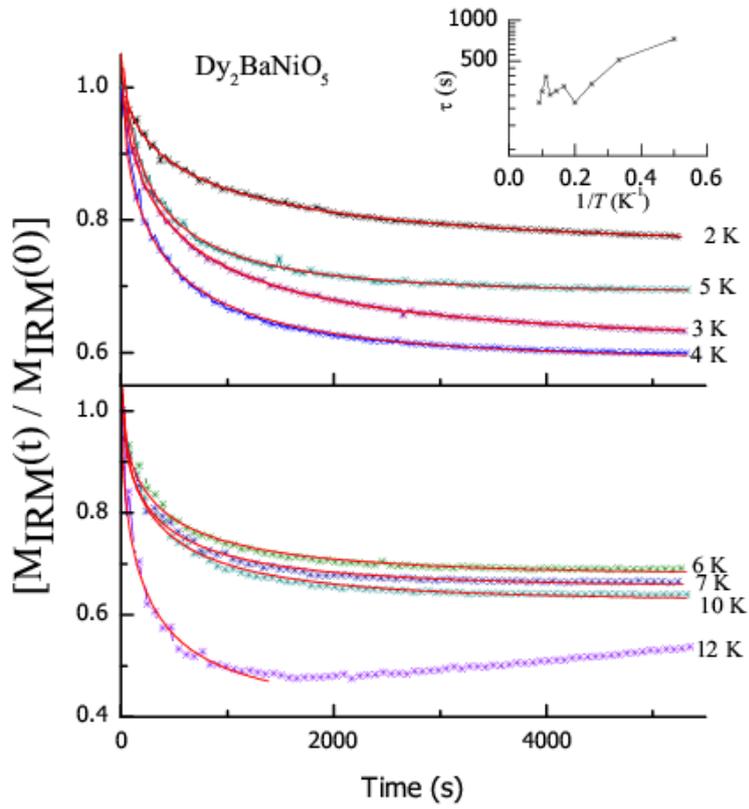

Figure 2:
Isothermal remnant magnetization measured as a function of time for several temperatures as described in the text. The continuous lines are the fits to a stretched exponential form of the type, $M_{IRM}(t)= A + B \exp[-t/\tau]^{0.5}$, where A and B constants and $\tau$ is the relaxation time. It is apparent that this functional form does not fit well for the temperatures mentioned in panel (b) and that there are qualitative changes in the shapes of the curves with varying temperature as demonstrated for 12 K (for which the fitting was carried below 1000 s only. Inset shows the plot of $\tau$ (in logarithmic scale) versus inverse temperature and the line through the data points is a guide to the eyes.

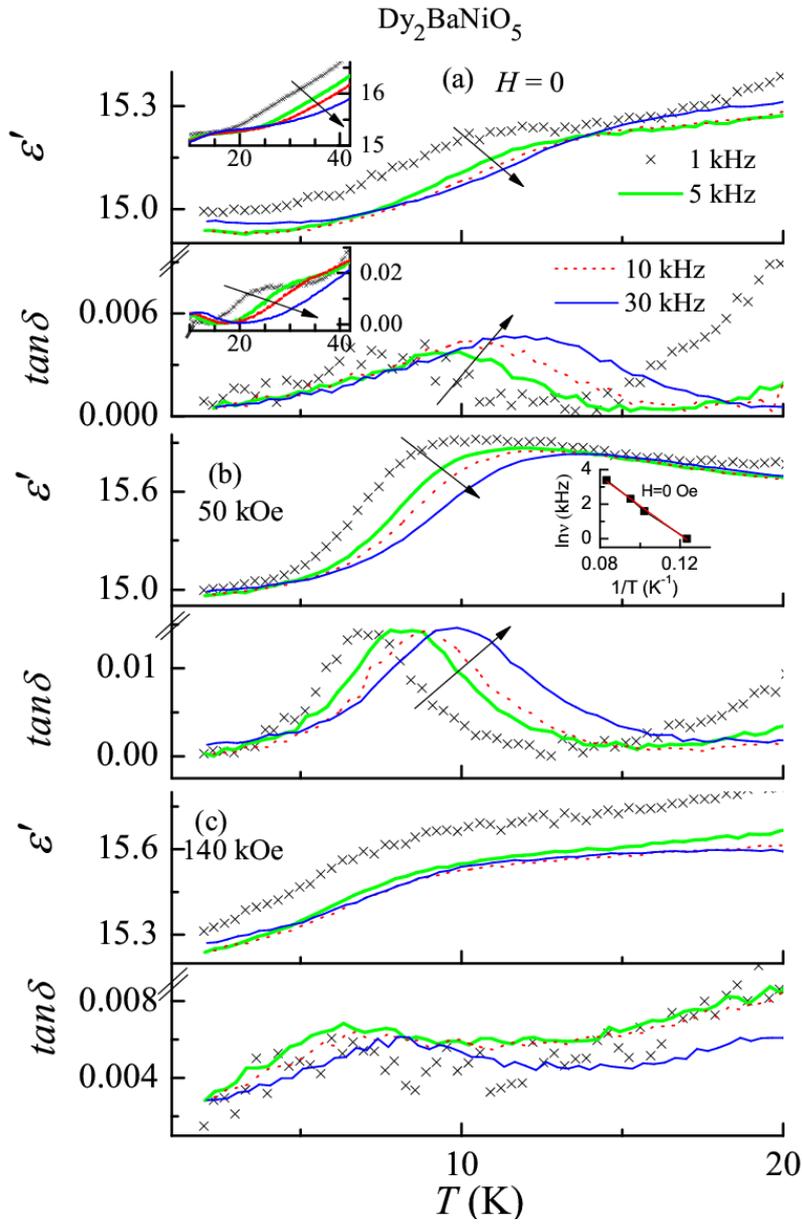

Figure 3:
Temperature dependence of dielectric constant ($\varepsilon'$) and loss factor ($\tan \delta$) obtained with various frequencies (a) in the absence of external magnetic-field (top two panels), (b) in 50 kOe (middle two panels) and (c) 140 kOe (bottom two panels). The arrows are shown to identify the curves with increasing frequency. The curves for 1 kHz is relatively more noisy. In the insets of (a), for zero-field, the curves are shown in the range 10-42 K. In the inset of (b), the inverse of the peak temperature in $\tan\delta(T)$ is plotted as a function of frequency (for the 10 K feature in the absence of $H$) to show the validity of Arrhenius behavior.

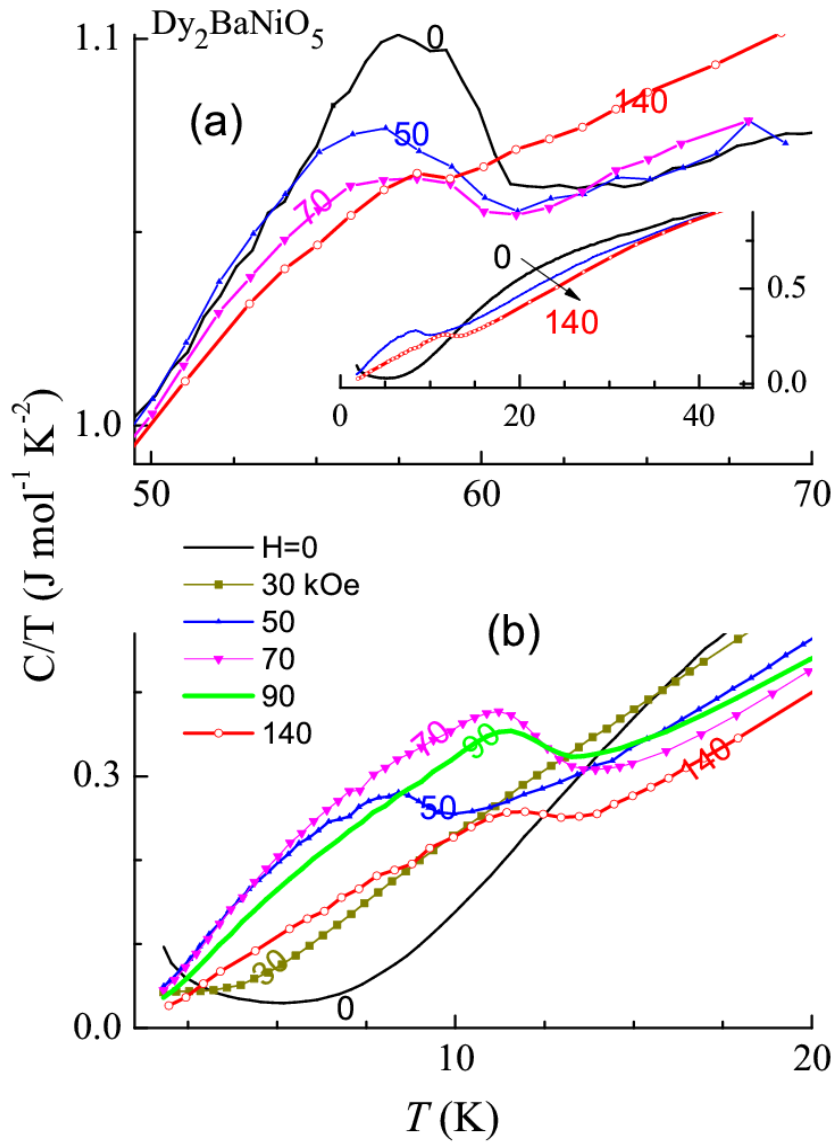

Figure 4:
Heat-capacity divided by temperature as a function of temperature in the range (a) 50 - 70 K and (b) 2- 20 K, measured in the presence of various magnetic fields. In the inset, the curves for the presence of some selected fields (0, 50 and 140 kOe) are shown with the arrow placed to show the way the curves move with increasing $H$ in that temperature region. The magnetic-field (in kOe) values are also placed on the curves for identification of curves.

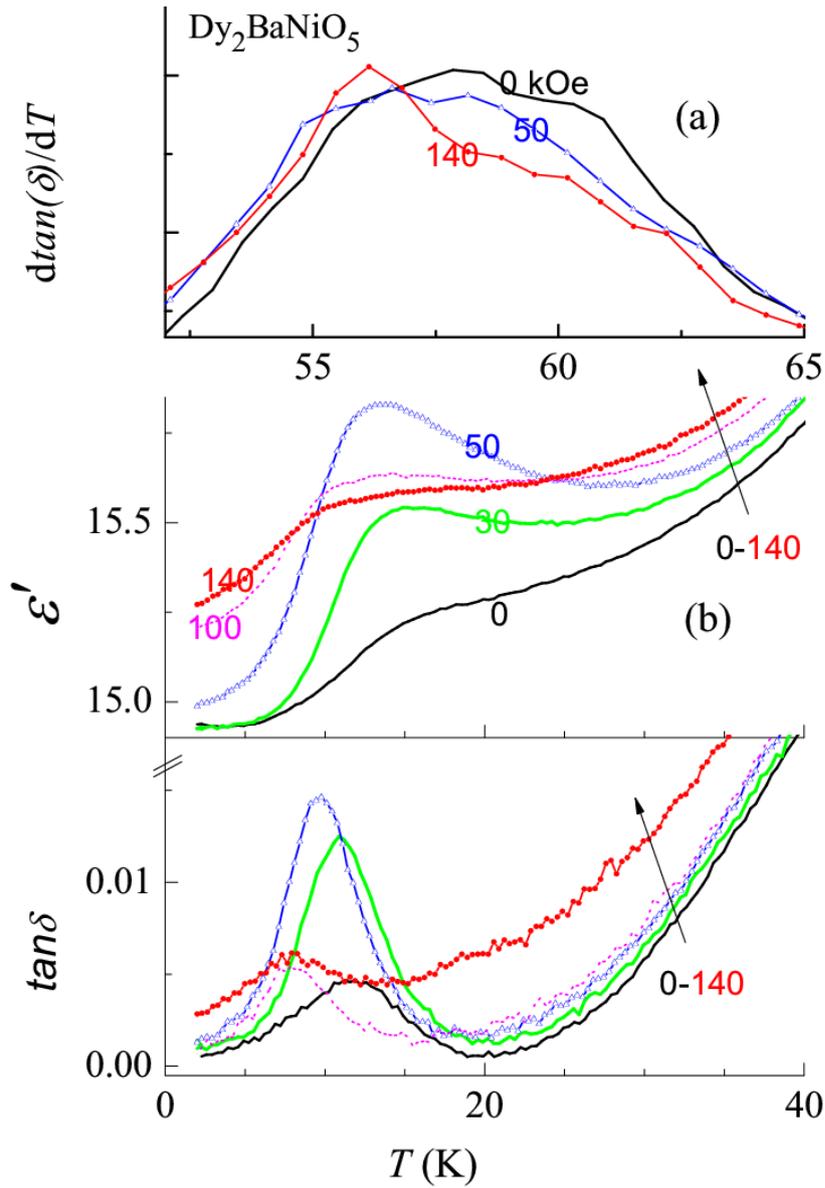

Figure 5:
(a) Temperature derivative of dielectric loss factor (*tan δ*) obtained with 30 kHz in the vicinity of the onset of magnetic order; (b) Temperature dependence of dielectric constant (*ε'*) and *tan δ* obtained with 30 kHz, in the presence of external magnetic-fields (0-140 kOe). The arrows serve as guides to identify the curves with increasing magnetic-field. Lines through data points only are shown.